\documentclass[conference]{worldcomp}
\IEEEoverridecommandlockouts
\usepackage[format=hang]{caption}
\usepackage{amsfonts,amsmath,amssymb}
\usepackage[table]{xcolor}
\usepackage[all]{xy}
\usepackage{graphicx}
\usepackage{fullpage}
\usepackage{setspace}
\usepackage{multirow}
\usepackage{longtable}
\usepackage{float}
\usepackage{subfig}
\usepackage[latin1]{inputenc}
\usepackage{tikz}
\usetikzlibrary{shapes,arrows}
\usepackage{pdfpages}
\usepackage{mathdots}
\usepackage{bbm}
\newcommand{\be}{\begin{equation}}
\newcommand{\ee}{\end{equation}}
\newcommand{\ben}{\begin{equation*}}
\newcommand{\een}{\end{equation*}}
\newcommand{\ZZ}{{\mathbbm{Z}}}
\newcommand{\NN}{{\mathbbm{N}}}

\newcommand{\card}{\mathop{\mathrm{card}}}
\newcommand{\f}{{\mathbf f}}

\newcommand{\B}{{\mathcal B}}

\newcommand{\G}{{\mathcal A}}
\newcommand{\N}{{\mathcal N}}

\newcommand{\calZ}{{\mathcal Z}}
\newcommand{\x}{{\vec{x}}}
\newcommand{\uu}{{\vec{u}}}
\newcommand{\vv}{{\vec{v}}}
\newcommand{\nn}{{\vec{n}}}
\newtheorem{prop}{Proposition}[section]

\newtheorem{thm}{Theorem}[section]

\newtheorem{procedure}{Procedure}[section]
\makeatletter
\let\@currsize\normalsize
\makeatother
\usepackage{flushend}

\tikzstyle{decision} = [rectangle, draw, fill=blue!20, 
    text width=12em, text centered, node distance=5cm, minimum height=4em]
    \tikzstyle{iterate} = [diamond, draw, fill=yellow!20, 
    text width=4em, text centered, node distance=5cm, minimum height=4em]
\tikzstyle{line} = [draw, -latex']
\tikzstyle{cloud} = [draw, ellipse,fill=red!20, node distance=3cm,
    minimum height=2em]
    \tikzstyle{cloudgood} = [draw, ellipse,fill=green!20, node distance=3cm,
    minimum height=2em]
        \tikzstyle{cloudunknown} = [draw, ellipse,fill=gray!20, node distance=3cm,
    minimum height=2em]

\begin{document}

\title{\bf Classification of two-dimensional binary \\ cellular automata with respect to surjectivity}           %%%% Replace with your title.

%%%% Replace the author and institution/affiliation names. 
%%%% Make sure the author names are boldface.
%\author{
%{\bfseries Henryk Fuk\'s and Andrew Skelton}\\
%Department of Mathematics, Brock University, St. Catharines, Ontario, Canada.
%}
%\title{Response Curves and Preimage Sequences of Two-Dimensional Cellular Automata}%\thanks{123}

% AUTHOR
\author{Henryk Fuk\'s and Andrew Skelton \\
Department of Mathematics\\
Brock University\\
St. Catharines, ON, Canada\\
}
\maketitle

\begin{abstract}
\noindent
While the surjectivity of the global map in two-dimensional cellular automata (2D CA) is undecidable in general, in specific cases one can often
decide if the rule is surjective or not. We attempt to classify as many 2D CA as possible by using a sequence of tests based on
the balance theorem, injectivity of the restriction to finite configurations, as well as permutivity. We introduce the notion of slice
permutivity which is shown to imply surjectivity in 2D CA. The tests are applied to 2D binary CA with neighbourhoods consisting of up to five sites,
considering all possible contiguous shapes of the neighbourhood. We find that if the size of the neighbourhood is less than five, 
complete classification of all rules is possible. Among 5-site rules, those with von Neuman neighbourhoods as well as neighbourhoods corresponding
to T, V, and Z pentominos can also be completely classified. 
\end{abstract}

\vspace{1em}

\section{Introduction}
 In the theory of cellular automata (CA), the surjectivity of the global map is one of the most extensively studied properties of CA. It is only natural to ask, therefore, what are the examples of surjective CA? 

In the case of one-dimensional CA, such examples are easy to construct because there exists  the well-known Amoroso-Patt algorithm for determining if a given elementary cellular automaton is surjective \cite{amoroso72}. Using this algorithm it can be shown that among the 88 minimal elementary CA rules, the only surjective rules have
Wolfram code numbers 15, 30, 45, 51, 60, 90, 105, 106, 150, 154, 170 and 204. 

In two dimensions, however, the situation is much different. It has been shown that the question of surjectivity of two-dimensional cellular automata is undecidable \cite{kari94}, which means that it is impossible to construct a single algorithm which would always decide if an
arbitrary rule is surjective or not. This, of course, does not exclude a possibility that for specific classes of 2D rules surjectivity can still be
decidable -- it is known, for example, that rules which are   permutive with respect to the corners of the Moore neighbourhood are surjective~\cite{dennunzio08}. 

In this paper, we attempt to classify 2D rules with respect to surjectivity using two known properties equivalent to surjectivity, namely the balance theorem and the injectivity of restrictions to finite configurations. Moreover, we introduce the the concept of slice-permutivity
which is then shown to imply surjectivity. We show that all 2D CA with neighbourhoods 
of size four (or less), no matter what shape, can be classified. For five-site CA, complete classification is still possible for certain
neighbourhood shapes, notably including von Neumann neighbourhood.

\section{Basic Definitions}

Let $\G$ be a finite set of symbols, to be called a {\em symbol set}. We define a {\em two-dimensional configuration} $s$ to be a function $s: \ZZ^2 \to \G$, and $\G^{\ZZ^2}$ to be the set of all two-dimensional configurations. For any vector $\x \in \ZZ^2$, we denote $s_\x \in \G$ to be a symbol located at position (or site) $\x$ in configuration $s$. If ${\cal V} \subset \ZZ^2$, we define $s_{\cal V}=[s_\x]_{\x \in \cal V}$. 

A {\em neighbourhood} $\N$ is a finite subset of vectors in $\ZZ^2$. A neighbourhood is said to be {\em contiguous} if, for any vector $\x \in \N$, at least one vector in the set $\{\x\pm (1,0), \x\pm(0,1)\}$ is also in $\N$. For any vector $\x$, the {\em neighbourhood of} $\x$ is defined as $\N(\x)=\{\uu+\x : \uu \in \N\}$.

We can now define the {\em local mapping} of a {\em two-dimensional cellular automata (2D CA)} to be the function $f: \G^{\N} \to \G$. The local mapping induces a {\em global mapping} $F : \G^{\ZZ^2} \to \G^{\ZZ^2}$ so that $F(s)_{\x} = f(s_{\N(\x)})$, for all $s \in \G^{\ZZ^2}$ and all $\x \in \ZZ^2$.

If $\calZ \subset \ZZ^2$ is a finite set of vectors, then we define a {\em block} to be an element
of $\G^\calZ$.

 The neighbourhood of $\calZ$ is defined similarly as before, so that $\N(\calZ) = \{\uu+\x : \uu \in \N, \x \in \calZ\}$. The {\em block evolution operator}  $\f : \G^{\N(\calZ)} \to \G^{\calZ}$ is now defined
by
$\f(b)_{\x}=f(b_{\N(\x)})$ for any $\x \in \calZ$ and $b \in \G^{\N(\calZ)}$.

Given a block $b \in \G^{\calZ}$, the set of preimages of $b$ under $\f$ is the set of blocks $b' \in
 \G^{\N(\calZ)}$ such that $\f(b') = b$. This preimage set will be denoted $\f^{-1}(b)$.

Sometimes, we will need to consider the neighbourhood of a neighbourhood. We will then use the notation $\N^2(\calZ) = \N(\N(\calZ))$,  and higher powers will refer to the appropriate number of neighbourhood compositions.

Let $\vv \in \N$, and let us denote ${\cal M}=\N \setminus \vv$.  Let $b \in \G^{\cal M}$ and let us denote
 $[x,b]$ to be an element of $\G^{\N}$ such that its entries with indices in $\cal M$ are the same as corresponding entries in $b$, while
the entry with index $\vv$ is equal to $x$, $x \in \G$.
A 2D CA is {\em permutive with respect to $\vv \in \N$} if, for any choice of $b$, the function
$x \rightarrow f([x,b])$ is one-to-one.

\subsection{One-dimensional Binary Rules}\label{results1d}

%%%%%%%%%%%
\begin{table*}
\centering
\begin{tabular}{c|c|c|c|c}
\multirow{2}{*}{Neighbourhood Size} & \multirow{2}{*}{Total Rules} &\multicolumn{3}{c}{Surjective Rules} \\ \cline{3-5}
& & Permutive & Not Permutive & Total \\ \hline\hline
1& 2    &2 &0&2 \\ \hline
2& 16  &6&0 &6\\ \hline
3& 256 &28&0 &28\\ \hline
4& 65536  &518&64 & 582 \\ \hline
5& 4294967296&131502&11516& 143018 \\ \hline
\end{tabular} 
\caption{One-dimensional binary rules}\label{table:1D}
\end{table*}

Before we attempt to classify two-dimensional CA rules, let us discuss what happens in one dimension,
as this will give us some important insight. As mentioned in the introduction, surjectivity in 1D is known to be
decidable, and the algorithm for testing for surjectivity has been developed by Amoroso and Patt 
in early 70's \cite{amoroso72}. We used this algorithm to find all surjective binary rules of a given neighbourhood size,
for neighbourhood sizes ranging from 1 to 5.  We also checked which of these rules are  permutive. The results are given in Table~\ref{table:1D}. 
One can make two interesting observations from this table. First of all,  the proportion of rules which are surjective decreases dramatically as the neighbourhood size increases. The second observation can be stated 
as the following proposition.

\vspace{5mm}

\begin{prop} \label{prop:1d}
Any contiguous one-dimensional binary cellular automata dependent on three or less sites is surjective if and only if it is permutive.
\end{prop}

\vspace{5mm}

This means that for a binary rule to be surjective yet non-permutive  a neighbourhood of at least four sites is needed.
A natural question to ask, therefore, is whether this is also the case in two dimensions?

\section{Permutivity and Surjectivity}
As we will shortly see, permutivity alone is not enough to guarantee surjectivity in two dimensions. 
In \cite{dennunzio08}, the authors considered 2D CA with Moore neighbourhood of radius $r$, where $\N = \{(i,j) : |i|,|j| \leq r\}$. They proved that any such rule is surjective if it is permutive with respect to sites $(\pm r,\pm r)$. We will prove a similar result using an arbitrary neighbourhood and any site that can be sliced off from the neighbourhood
by a straight line.

Given $m,c \in \mathbb{Q}$, we define a line $\ell = \{(x,y) :  y = mx+c\}$, and the following two regions, $\ell^+ = \{(x,y) : y > mx+c\}$ and $\ell^- = \{(x,y) : y < mx+c\}$. For vertical lines $\ell = \{(x,y) :  x=c\}$ we similarly define $\ell^+ = \{(x,y) : x > c\}$ and $\ell^- = \{(x,y) : x < c\}$.
 A site $\x \in \N$ can be {\em sliced} if there exists a set $\ell$ such that $\x \in \ell$ and $\N \setminus \x \subset \ell^+ \; (\text{or } \ell^-)$. A 2D CA is {\em slice permutive} if it is permutive with respect to a site which can be sliced.

The main result relating slice-permutivity and surjectivity can be stated as follows.

\vspace{5mm}

\begin{thm} \label{thm:slicepermutive}
Any two-dimensional slice permutive CA is surjective.
\end{thm}

\vspace{5mm}

Before we start the proof, we will need the following classical result. Let $\calZ_n$ be a square region
in $\ZZ^2$, defined as $\calZ_n=[0,1,\ldots, n-1] \times [0,1,\ldots, n-1]$, where $ n \in \NN$.

\vspace{5mm}

\begin{thm}[Balance Theorem] \label{thm:balance}
A 2D CA is surjective if and only if for all $n \geq 1$ and all $b,b' \in \G^{\calZ_n}$, we have $\card \f^{-1}(b) = \card \f^{-1}(b')$.
\end{thm}

\vspace{5mm}

A one-dimensional version of this theorem first appeared in \cite{hedlund69}. 
The proof of the two-dimensional version can be found in \cite{maruoka76}, where the authors consider a Moore neighbourhood of any radius. Since any neighbourhood can be extended to a Moore neighbourhood by adding extra sites, the Balance Theorem also holds for a CA rule with any neighbourhood shape.

\vspace{1.5mm}

\begin{proof}[of Theorem \ref{thm:slicepermutive}]
Consider an arbitrary block $b \in \G^{\calZ_n}$. Since our CA is slice permutive, there exists a line $\ell$ which slices the neighbourhood at some site $\uu \in \N$, so that all sites in $\N \setminus \uu$ are either in $\ell^-$ or in $\ell^+$. Without loss of generality, we assume that
all sites of $\N \setminus \uu$ are  are in $\ell^-$. The proof of the other case can be obtained by replacing $\ell^\pm$ with $\ell^\mp$. Let $\nn$ be the normal vector to $\ell$ oriented so that it points in the direction of $\ell^+$. Consider $\{\ell_1, \ell_2, \dots, \ell_k\}$ to be the family of lines with the same slope as $\ell$ with indices increasing in the direction of $\nn$, so that for any $\x \in \calZ_n$ there exists $\ell_i$ such that $\x \in \ell_i$. We now construct the set of preimages of $b$, each element of which is of the form $[a_{\x}]_{\x \in \N(\calZ_n)}$.

In order to illustrate this better, we will conduct the proof while simultaneously referring to an example of
 a rule defined on the seven-site neighbourhood 
\begin{align*}
\N&=\{(-3,1),(-3,0),(-2,0),(-1,0),\\&\quad\quad\quad(0,0),(0,-1),(1,-1)\}.
\end{align*}
In this example, we assume that the  rule is permutive with respect to the sliceable site $(0,0)$, for which the corresponding line $\ell$ has slope $-1/2$. The neighbourhood of $\calZ_3$ is shown in Figure \ref{fig:procedure}(a).

We start the construction from sites of $\calZ_n$ which belong to the line $l_1$. Due to slice permutivity, all sites of the preimage
which are below this line can take arbitrary values, and sites which are on the line can be chosen in such a way that
$b_\x=f(a_{\N(\x)})$ for all $\x \in \ell_1 \cap \calZ_n$. This means that  sites of $\N(\calZ_n)$ which belong to $\ell_1^-$ can take arbitrary values,
and sites of $\calZ_n$ which belong to $\ell_1$ are uniquely determined by those arbitrary values. 

In Figure \ref{fig:procedure}(b)
arbitrary sites are denoted by stars, and the uniquely determined site (only one in this case) is denoted by $\textrm D$.

We then move to $\ell_2$, and again, for every possible configuration of sites from the set $\N(\calZ_n) \cap \ell_2^-$, values
of sites which lie in $\calZ_n \cap \ell_2$ will be uniquely determined. It may happen, as shown in Figure \ref{fig:procedure}(b), that
some sites of $\N(\calZ_n) \cap \ell_2^-$ have not been labeled before.   These sites can also assume arbitrary values,
and thus they are marked as red stars in Figure \ref{fig:procedure}(c).

We repeat the above procedure  until all sites of $\N(\calZ_n)$ are labeled, as shown in Figures \ref{fig:procedure}(d-h).
In the end, as in Figure \ref{fig:procedure}(i), all sites which belong to $\calZ_n$ will be labeled by $D$, while the
sites of $\N(\calZ_n) \setminus \calZ_n$ will be labeled by stars. This means that we have $\card (\N(\calZ_n) \setminus \calZ_n)$
sites in the preimage which can assume any values from the symbol set $\G$.
% ORIGINAL
% For $i=1$ to $i=k$, consider options for all undetermined values of $a_{\x}$ as follows:
% \begin{itemize}
% \item If $i \geq 2$ and $\x \in \ell_{i-1}^-$, then the value of $a_{\x}$ has either already been assigned a fixed symbol, or has already been arbitrarily chosen.
% \item If $\x \in \ell_{i}^-$, then $a_{\x}$ can be arbitrarily chosen.
% \item If $\x \in \ell_i \cap \calZ_n$, then $a_{\x}$ has exactly one fixed symbol for each possible value of $b_{\x}$ and each previously considered and arbitrarily assigned option for $\N(a_{\x}) \setminus a_{\x}$. This is due to the definition and assumption of slice permutivity.
% \item If $\x \in \ell_i$ and $\x \notin \calZ_n$, then no action is taken.
% \end{itemize}
% For all $\x \in \calZ_n$, there is a single value of $a_{\x}$ forced by the procedure. For all $\x \in \N(\calZ_n) \setminus \calZ_n$, the value of $a_{\x}$ is arbitrarily chosen. 
Thus, for each block $b \in \B_r$, we have $\card \f^{-1}(b) = (\card \G)^{\card (\N(\calZ_n) \setminus \calZ_n)}$. Since this is independent of $b$, by Theorem \ref{thm:balance} we conclude that the CA is surjective.
\end{proof}
% \begin{example}
% Consider a rule defined on the seven-site neighbourhood 
% $$\N=\{(-3,1),(-3,0),(-2,0),(-1,0),(0,0),(0,-1),(1,-1)\}.$$ 
% Assume that our cellular automata rule is permutive with respect to the sliceable site $(0,0)$, for which the corresponding line $\ell$ has slope $-1/2$. In Figure \ref{fig:procedure}, we use the procedure of Theorem \ref{thm:slicepermutive} to find the set of preimages of a block $b \in \B_3$. We know that $|\N(\calZ_3)|= 27$ (see Figure \ref{fig:step0} where the sites in $\calZ_3$ have been highlighted). In each step (Figures \ref{fig:step1}-\ref{fig:step7}), an $F$ denotes a fixed site symbol and $\star$ denotes an arbitrarily chosen site symbol. In Figure \ref{fig:step8}, we see that for any $b \in \G^{\calZ_3}$, we have $\card \f^{-1}(b) = N^{|\N(\calZ_n) \setminus \calZ_n|} = (\card \G)^{18}$.
% \end{example}

\begin{figure*}
\centering
\subfloat[$\N(Z_3)$]{\label{fig:step0}
\begin{tikzpicture}
\draw[ densely dashed] (-1.5, 0) circle (3pt);
\draw[ densely dashed] (-1.5, 0.5) circle (3pt);
\draw[ densely dashed] (-1.5, 1) circle (3pt);
\draw[ densely dashed] (-1.5, 1.5) circle (3pt);
\draw[ densely dashed] (-1, 0) circle (3pt);
\draw[ densely dashed] (-1, 0.5) circle (3pt);
\draw[ densely dashed] (-1, 1) circle (3pt);
\draw[ densely dashed] (-1, 1.5) circle (3pt);
\draw[ densely dashed] (-0.5, 0) circle (3pt);
\draw[ densely dashed] (-0.5, 0.5) circle (3pt);
\draw[ densely dashed] (-0.5, 1) circle (3pt);
\draw[ densely dashed] (-0.5, 1.5) circle (3pt);
\draw[ solid, thick] (0, 0) circle (3pt); 
\draw[ solid, thick] (0, 0.5) circle (3pt);
\draw[ solid, thick] (0, 1) circle (3pt);
\draw[ densely dashed] (0, -0.5) circle (3pt);
\draw[ densely dashed] (0.5, -0.5) circle (3pt);
\draw[ solid, thick] (0.5, 0) circle (3pt);
\draw[ solid, thick] (0.5, 0.5) circle (3pt);
\draw[ solid, thick] (0.5, 1) circle (3pt);
\draw[ densely dashed] (1, -0.5) circle (3pt);
\draw[ solid, thick] (1, 0) circle (3pt);
\draw[ solid, thick] (1, 0.5) circle (3pt);
\draw[ solid, thick] (1, 1) circle (3pt);
\draw[ densely dashed] (1.5, 0) circle (3pt);
\draw[ densely dashed] (1.5, 0.5) circle (3pt);
\draw[ densely dashed] (1.5, -0.5) circle (3pt);
\draw[white] (2.2,-1) node{$\ell_0$};
\end{tikzpicture}}\hspace{5mm}
\subfloat[$i=1$]{\label{fig:step1}
\begin{tikzpicture}
\draw[ densely dashed] (-1.5, 1) circle (3pt);
\draw[ densely dashed] (-1.5, 1.5) circle (3pt);
\draw[ densely dashed] (-1, 0.5) circle (3pt);
\draw[ densely dashed] (-1, 1) circle (3pt);
\draw[ densely dashed] (-1, 1.5) circle (3pt);
\draw[ densely dashed] (-0.5, 0.5) circle (3pt);
\draw[ densely dashed] (-0.5, 1) circle (3pt);
\draw[ densely dashed] (-0.5, 1.5) circle (3pt);
\draw[ densely dashed] (0, 0.5) circle (3pt);
\draw[ densely dashed] (0, 1) circle (3pt);
\draw[ densely dashed] (0.5, 0) circle (3pt);
\draw[ densely dashed] (0.5, 0.5) circle (3pt);
\draw[ densely dashed] (0.5, 1) circle (3pt);
\draw[ densely dashed] (1, -0.5) circle (3pt);
\draw[ densely dashed] (1, 0) circle (3pt);
\draw[ densely dashed] (1, 0.5) circle (3pt);
\draw[ densely dashed] (1, 1) circle (3pt);
\draw[ densely dashed] (1.5, 0) circle (3pt);
\draw[ densely dashed] (1.5, 0.5) circle (3pt);
\draw[ densely dashed] (1.5, -0.5) circle (3pt);
\draw (-2,1) -- (2,-1); \draw (2.2,-1) node{$\ell_1$};
\draw (-1,0)[red] node{$\star$};
\draw (-0.5,0)[red] node{$\star$};
\draw (-1.5,0)[red] node{$\star$};
\draw (-1.5,0.5)[red] node{$\star$};
\draw (0,-0.5)[red] node{$\star$};
\draw (0.5,-0.5)[red] node{$\star$};
\draw (0,0)[blue] node{D};
\end{tikzpicture}}\hspace{5mm}
\subfloat[$i=2$]{\label{fig:step2}
\begin{tikzpicture}
\draw[ densely dashed] (-1.5, 1) circle (3pt);
\draw[ densely dashed] (-1.5, 1.5) circle (3pt);
\draw[ densely dashed] (-1, 1) circle (3pt);
\draw[ densely dashed] (-1, 1.5) circle (3pt);
\draw[ densely dashed] (-0.5, 0.5) circle (3pt);
\draw[ densely dashed] (-0.5, 1) circle (3pt);
\draw[ densely dashed] (-0.5, 1.5) circle (3pt);
\draw[ densely dashed] (0, 0.5) circle (3pt);
\draw[ densely dashed] (0, 1) circle (3pt);
\draw[ densely dashed] (0.5, 0.5) circle (3pt);
\draw[ densely dashed] (0.5, 1) circle (3pt);
\draw[ densely dashed] (1, 0) circle (3pt);
\draw[ densely dashed] (1, 0.5) circle (3pt);
\draw[ densely dashed] (1, 1) circle (3pt);
\draw[ densely dashed] (1.5, 0) circle (3pt);
\draw[ densely dashed] (1.5, 0.5) circle (3pt);
\draw[ densely dashed] (1.5, -0.5) circle (3pt);
\draw[gray!50!] (-2,1) -- (2,-1); %\draw[gray!50!] (2.2,-1.2) node{$\ell_0$};
\draw (-2,1.25) -- (2,-0.75); \draw (2.2,-0.75) node{$\ell_2$};
\draw (-1,0)[] node{$\star$};
\draw (-0.5,0)[] node{$\star$};
\draw (-1.5,0)[] node{$\star$};
\draw (-1.5,0.5)[] node{$\star$};
\draw (0,-0.5)[] node{$\star$};
\draw (0.5,-0.5)[] node{$\star$};
\draw (0,0)[] node{D};
\draw (-1,0.5)[red] node{$\star$};
\draw (1,-0.5)[red] node{$\star$};
\draw (0.5,0)[blue] node{D};
\end{tikzpicture}}
\vspace{5mm}

\subfloat[$i=3$]{\label{fig:step3}
\begin{tikzpicture}
\draw[ densely dashed] (-1.5, 1.5) circle (3pt);
\draw[ densely dashed] (-1, 1) circle (3pt);
\draw[ densely dashed] (-1, 1.5) circle (3pt);
\draw[ densely dashed] (-0.5, 1) circle (3pt);
\draw[ densely dashed] (-0.5, 1.5) circle (3pt);
\draw[ densely dashed] (0, 1) circle (3pt);
\draw[ densely dashed] (0.5, 0.5) circle (3pt);
\draw[ densely dashed] (0.5, 1) circle (3pt);
\draw[ densely dashed] (1, 0.5) circle (3pt);
\draw[ densely dashed] (1, 1) circle (3pt);
\draw[ densely dashed] (1.5, 0) circle (3pt);
\draw[ densely dashed] (1.5, 0.5) circle (3pt);
\draw[gray!50!] (-2,1) -- (2,-1); %\draw[gray!50!] (2.2,-1.2) node{$\ell_0$};
\draw[gray!50!] (-2,1.25) -- (2,-0.75); %\draw (2.2,-0.75) node{$\ell_1$};
\draw (-2,1.5) -- (2,-0.5); \draw (2.2,-0.5) node{$\ell_3$};
\draw (-1,0)[] node{$\star$};
\draw (-0.5,0)[] node{$\star$};
\draw (-1.5,0)[] node{$\star$};
\draw (-1.5,0.5)[] node{$\star$};
\draw (0,-0.5)[] node{$\star$};
\draw (0.5,-0.5)[] node{$\star$};
\draw (0,0)[] node{D};
\draw (-1,0.5)[] node{$\star$};
\draw (1,-0.5)[] node{$\star$};
\draw (0.5,0)[] node{D};
\draw (-0.5,0.5)[red] node{$\star$};
\draw (-1.5,1)[red] node{$\star$};
\draw (1.5,-0.5)[red] node{$\star$};
\draw (0,0.5)[blue] node{D};
\draw (1,0)[blue] node{D};
\end{tikzpicture}} \hspace{5mm}
\subfloat[$i=4$]{\label{fig:step4}
\begin{tikzpicture}
\draw[ densely dashed] (-1.5, 1.5) circle (3pt);
\draw[ densely dashed] (-1, 1.5) circle (3pt);
\draw[ densely dashed] (-0.5, 1) circle (3pt);
\draw[ densely dashed] (-0.5, 1.5) circle (3pt);
\draw[ densely dashed] (0, 1) circle (3pt);
\draw[ densely dashed] (0.5, 1) circle (3pt);
\draw[ densely dashed] (1, 0.5) circle (3pt);
\draw[ densely dashed] (1, 1) circle (3pt);
\draw[ densely dashed] (1.5, 0) circle (3pt);
\draw[ densely dashed] (1.5, 0.5) circle (3pt);
\draw[gray!50!] (-2,1) -- (2,-1); %\draw[gray!50!] (2.2,-1.2) node{$\ell_0$};
\draw[gray!50!] (-2,1.25) -- (2,-0.75); %\draw (2.2,-0.75) node{$\ell_1$};
\draw[gray!50!] (-2,1.5) -- (2,-0.5); %\draw (2.2,-0.5) node{$\ell_2$};
\draw (-2,1.75) -- (2,-0.25); \draw (2.2,-0.25) node{$\ell_4$};
\draw (-1,0)[] node{$\star$};
\draw (-0.5,0)[] node{$\star$};
\draw (-1.5,0)[] node{$\star$};
\draw (-1.5,0.5)[] node{$\star$};
\draw (0,-0.5)[] node{$\star$};
\draw (0.5,-0.5)[] node{$\star$};
\draw (0,0)[] node{D};
\draw (-1,0.5)[] node{$\star$};
\draw (1,-0.5)[] node{$\star$};
\draw (0.5,0)[] node{D};
\draw (-0.5,0.5)[] node{$\star$};
\draw (-1.5,1)[] node{$\star$};
\draw (1.5,-0.5)[] node{$\star$};
\draw (0,0.5)[] node{D};
\draw (1,0)[] node{D};
\draw (-1,1)[red] node{$\star$};
\draw (0.5,0.5)[blue] node{D};
\end{tikzpicture}}\hspace{5mm}
\subfloat[$i=5$]{\label{fig:step5}
\begin{tikzpicture}
\draw[ densely dashed] (-1, 1.5) circle (3pt);
\draw[ densely dashed] (-0.5, 1.5) circle (3pt);
\draw[ densely dashed] (0.5, 1) circle (3pt);
\draw[ densely dashed] (1, 1) circle (3pt);
\draw[ densely dashed] (1.5, 0.5) circle (3pt);
\draw[gray!50!] (-2,1) -- (2,-1); %\draw[gray!50!] (2.2,-1.2) node{$\ell_0$};
\draw[gray!50!] (-2,1.25) -- (2,-0.75); %\draw (2.2,-0.75) node{$\ell_1$};
\draw[gray!50!] (-2,1.5) -- (2,-0.5); %\draw (2.2,-0.5) node{$\ell_2$};
\draw[gray!50!]  (-2,1.75) -- (2,-0.25); %\draw (2.2,-0.25) node{$\ell_3$};
\draw (-2,2) -- (2,0); \draw (2.2,0) node{$\ell_5$};
\draw (-1,0)[] node{$\star$};
\draw (-0.5,0)[] node{$\star$};
\draw (-1.5,0)[] node{$\star$};
\draw (-1.5,0.5)[] node{$\star$};
\draw (0,-0.5)[] node{$\star$};
\draw (0.5,-0.5)[] node{$\star$};
\draw (0,0)[] node{D};
\draw (-1,0.5)[] node{$\star$};
\draw (1,-0.5)[] node{$\star$};
\draw (0.5,0)[] node{D};
\draw (-0.5,0.5)[] node{$\star$};
\draw (-1.5,1)[] node{$\star$};
\draw (1.5,-0.5)[] node{$\star$};
\draw (0,0.5)[] node{D};
\draw (1,0)[] node{D};
\draw (-1,1)[] node{$\star$};
\draw (0.5,0.5)[] node{D};
\draw (-0.5,1)[red] node{$\star$};
\draw (-1.5,1.5)[red] node{$\star$};
\draw (1.5,0)[red] node{$\star$};
\draw (0,1)[blue] node{D};
\draw (1,0.5)[blue] node{D};
\end{tikzpicture}}
\vspace{5mm}

\subfloat[$i=6$]{\label{fig:step6}
\begin{tikzpicture}
\draw[ densely dashed] (-0.5, 1.5) circle (3pt);
\draw[ densely dashed] (1, 1) circle (3pt);
\draw[ densely dashed] (1.5, 0.5) circle (3pt);
\draw[gray!50!] (-2,1) -- (2,-1); %\draw[gray!50!] (2.2,-1.2) node{$\ell_0$};
\draw[gray!50!] (-2,1.25) -- (2,-0.75); %\draw (2.2,-0.75) node{$\ell_1$};
\draw[gray!50!] (-2,1.5) -- (2,-0.5); %\draw (2.2,-0.5) node{$\ell_2$};
\draw[gray!50!]  (-2,1.75) -- (2,-0.25); %\draw (2.2,-0.25) node{$\ell_3$};
\draw[gray!50!] (-2,2) -- (2,0); %\draw (2.2,0) node{$\ell_5$};
\draw (-2,2.25) -- (2,0.25); \draw (2.2,0.25) node{$\ell_6$};
\draw (-1,0)[] node{$\star$};
\draw (-0.5,0)[] node{$\star$};
\draw (-1.5,0)[] node{$\star$};
\draw (-1.5,0.5)[] node{$\star$};
\draw (0,-0.5)[] node{$\star$};
\draw (0.5,-0.5)[] node{$\star$};
\draw (0,0)[] node{D};
\draw (-1,0.5)[] node{$\star$};
\draw (1,-0.5)[] node{$\star$};
\draw (0.5,0)[] node{D};
\draw (-0.5,0.5)[] node{$\star$};
\draw (-1.5,1)[] node{$\star$};
\draw (1.5,-0.5)[] node{$\star$};
\draw (0,0.5)[] node{D};
\draw (1,0)[] node{D};
\draw (-1,1)[] node{$\star$};
\draw (0.5,0.5)[] node{D};
\draw (-0.5,1)[] node{$\star$};
\draw (-1.5,1.5)[] node{$\star$};
\draw (1.5,0)[] node{$\star$};
\draw (0,1)[] node{D};
\draw (1,0.5)[] node{D};
\draw (-1,1.5)[red] node{$\star$};
\draw (0.5,1)[blue] node{D};
\end{tikzpicture}} \hspace{5mm}
\subfloat[$i=7$]{\label{fig:step7}
\begin{tikzpicture}
\draw[gray!50!] (-2,1) -- (2,-1); %\draw[gray!50!] (2.2,-1.2) node{$\ell_0$};
\draw[gray!50!] (-2,1.25) -- (2,-0.75); %\draw (2.2,-0.75) node{$\ell_1$};
\draw[gray!50!] (-2,1.5) -- (2,-0.5); %\draw (2.2,-0.5) node{$\ell_2$};
\draw[gray!50!]  (-2,1.75) -- (2,-0.25); %\draw (2.2,-0.25) node{$\ell_3$};
\draw[gray!50!] (-2,2) -- (2,0); %\draw (2.2,0) node{$\ell_5$};
\draw[gray!50!] (-2,2.25) -- (2,0.25); %\draw (2.2,0.25) node{$\ell_6$};
\draw (-2,2.5) -- (2,0.5); \draw (2.2,0.5) node{$\ell_7$};
\draw (-1,0)[] node{$\star$};
\draw (-0.5,0)[] node{$\star$};
\draw (-1.5,0)[] node{$\star$};
\draw (-1.5,0.5)[] node{$\star$};
\draw (0,-0.5)[] node{$\star$};
\draw (0.5,-0.5)[] node{$\star$};
\draw (0,0)[] node{D};
\draw (-1,0.5)[] node{$\star$};
\draw (1,-0.5)[] node{$\star$};
\draw (0.5,0)[] node{D};
\draw (-0.5,0.5)[] node{$\star$};
\draw (-1.5,1)[] node{$\star$};
\draw (1.5,-0.5)[] node{$\star$};
\draw (0,0.5)[] node{D};
\draw (1,0)[] node{D};
\draw (-1,1)[] node{$\star$};
\draw (0.5,0.5)[] node{D};
\draw (-0.5,1)[] node{$\star$};
\draw (-1.5,1.5)[] node{$\star$};
\draw (1.5,0)[] node{$\star$};
\draw (0,1)[] node{D};
\draw (1,0.5)[] node{D};
\draw (-1,1.5)[] node{$\star$};
\draw (0.5,1)[] node{D};
\draw (-0.5,1.5)[red] node{$\star$};
\draw (1.5,0.5)[red] node{$\star$};
\draw (1,1)[blue] node{D};
\end{tikzpicture}}\hspace{5mm}
\subfloat[Final result]{\label{fig:step8}
\begin{tikzpicture}
\draw[gray!50!] (-2,1) -- (2,-1); %\draw[gray!50!] (2.2,-1.2) node{$\ell_0$};
\draw[gray!50!] (-2,1.25) -- (2,-0.75); %\draw (2.2,-0.75) node{$\ell_1$};
\draw[gray!50!] (-2,1.5) -- (2,-0.5); %\draw (2.2,-0.5) node{$\ell_2$};
\draw[gray!50!]  (-2,1.75) -- (2,-0.25); %\draw (2.2,-0.25) node{$\ell_3$};
\draw[gray!50!] (-2,2) -- (2,0); %\draw (2.2,0) node{$\ell_5$};
\draw[gray!50!] (-2,2.25) -- (2,0.25); %\draw (2.2,0.25) node{$\ell_6$};
\draw[gray!50!] (-2,2.5) -- (2,0.5); %\draw (2.2,0.5) node{$\ell_7$};
\draw (-1,0)[] node{$\star$};
\draw (-0.5,0)[] node{$\star$};
\draw (-1.5,0)[] node{$\star$};
\draw (-1.5,0.5)[] node{$\star$};
\draw (0,-0.5)[] node{$\star$};
\draw (0.5,-0.5)[] node{$\star$};
\draw (0,0)[] node{D};
\draw (-1,0.5)[] node{$\star$};
\draw (1,-0.5)[] node{$\star$};
\draw (0.5,0)[] node{D};
\draw (-0.5,0.5)[] node{$\star$};
\draw (-1.5,1)[] node{$\star$};
\draw (1.5,-0.5)[] node{$\star$};
\draw (0,0.5)[] node{D};
\draw (1,0)[] node{D};
\draw (-1,1)[] node{$\star$};
\draw (0.5,0.5)[] node{D};
\draw (-0.5,1)[] node{$\star$};
\draw (-1.5,1.5)[] node{$\star$};
\draw (1.5,0)[] node{$\star$};
\draw (0,1)[] node{D};
\draw (1,0.5)[] node{D};
\draw (-1,1.5)[] node{$\star$};
\draw (0.5,1)[] node{D};
\draw (-0.5,1.5)[] node{$\star$};
\draw (1.5,0.5)[] node{$\star$};
\draw (1,1)[] node{D};
\end{tikzpicture}}
\caption{Example: Iteration of the procedure in Theorem \ref{thm:slicepermutive}}
\end{figure*}
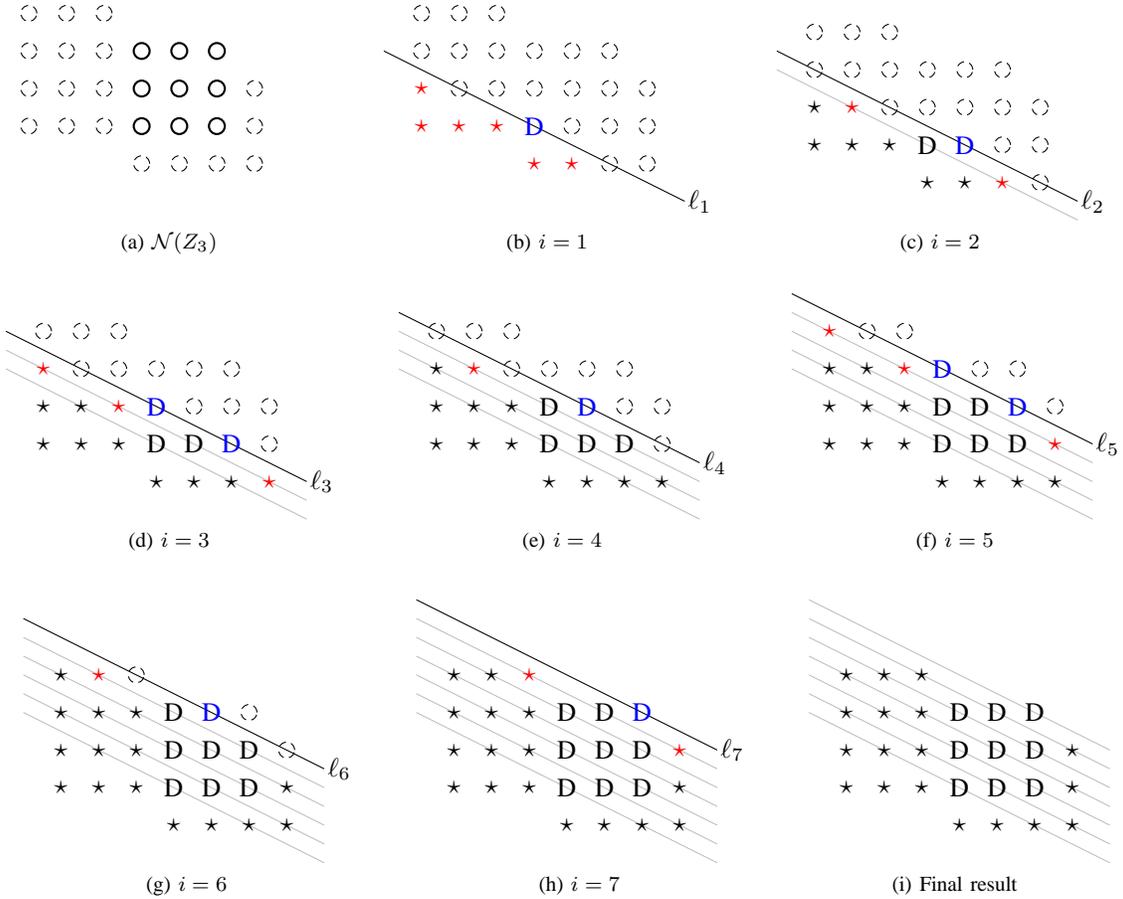 \label{fig:procedure}

\section{Classification procedure}
The result of the previous section gives us a method to determine surjectivity of a special class of
rules, namely slice-permutive rules. In classifying rules we will also need some method for checking if
a rule is not surjective -- although of  course a {\em general} algorithm of this type cannot exist.
One such method involves the Balance Theorem. One can simply check if all blocks of a given size
have the same number of preimages - if one finds a single violation of the balance theorem, the rule is obviously not surjective. Clearly, one can perform such exhaustive check only for block of small size, and if no balance violation is found, then the test remains inconclusive. It turns out, however, that this simple method is surprisingly effective
for 2D rules with small neighbourhood size.

Another test for non-surjectivity is based on the following classical result \cite{richardson72}. Here by a finite configuration we mean a configuration where all but a finite number of sites are in some arbitrary fixed state.

\vspace{5mm}

\begin{thm}  \label{thm:richardson}
A CA is surjective if and only if it is injective when restricted to finite configurations.
\end{thm} 

\vspace{5mm}

If one can find two finite configurations which are different but have the same image, then the rule is not surjective. One can perform this test exhaustively for small finite configurations. It turns out to be useful in cases where the balance theorem test is inconclusive.

We may now present a procedure for classification of 2D rules with a given
neighbourhood shape. This procedure is  illustrated in Figure~\ref{fig:flowchart}. 
While it has been tailored for binary rules (that is, rules where $\card \G=2$), it can be easily adapted for larger alphabets.

\vspace{5mm}

\begin{procedure} \label{proc:surjectivity}
We start with a list of all  rules on a given neighbourhood and apply the following filters to remove known surjective and non-surjective rules. 
\begin{enumerate}
\item First we check the simplest balance condition, that is, $\card \f^{-1}(0) = \card \f^{-1}(1)$. Rules which violate
this condition are non-surjective.
\item We check if the rule truly depended on all given sites. If not, it means it can be treated as if it had smaller neighbourhood, thus we check how it was classified
before (we were classifying rules with progressively increasing neighbourhood size). 
\item We check  if the rule is permutive with respect to at least one sliceable site. If it is, it is
surjective.
\item The following two tests are performed simultaneously owing to the fact that they required approximately the same information.
\begin{itemize}
\item We check if all  blocks from the set $\G^{\N(\calZ_1)}$ have the same number of preimages. If not, the rule is non-surjective.

\item We search for violations of injectivity on finite configurations. Let $\cal M$ be the set of all sites $\x \in
\ZZ^2$ for which $(0,0) \in \N(\x)$. For each $b \in \G^{\cal M}$ we consider a finite  configuration $s$ such that
$s_{\cal M}=b$ and the sites which do not belong to $\cal M$ are in the  state 0, as well as
configuration $s^\prime$ obtained from $s$ by replacing $s_{(0,0)}$ by $1-s_{(0,0)}$. If
$F(s) = F(s^\prime)$, injectivity is violated, thus the rule is not surjective.
\end{itemize}
\item We check if all  blocks from the set $\G^{\N^2(\calZ_1)}$ have the same number of preimages. If not, the rule is non-surjective.
\end{enumerate}
\end{procedure}

\begin{figure*}
\begin{centering}
\begin{tikzpicture}[node distance = 2.5cm, auto]
% Place nodes
\node [decision] (init) {For each rule with neighbourhood $\N$};
\node [decision, below of=init, node distance = 2.5cm] (ruletable) {Local function balanced?};
\node [cloud, right of =ruletable, node distance = 7cm] (non1) {Non-surjective};
\node [decision, below of = ruletable, node distance = 2.5cm] (depend) {Truly dependent on neighbourhood?};
\node [cloudunknown, right of =depend, node distance = 7cm] (non2) {Check previous results};
\node [decision, below of = depend, node distance = 2.5cm] (slice) {Slice permutive?};
\node [cloudgood, right of =slice, node distance = 7cm] (good1) {Surjective};
\node (myhillbalance1) [decision,rectangle split, rectangle split parts=2, node distance = 3cm, below of=slice, minimum height = 8em]{$\G^{\N^2(\calZ_1)} \to \G^{\N(\calZ_1)}$\\ balance violated?\nodepart{second} Fails injectivity test \\ on finite configurations?};
\node [decision, below of = myhillbalance1, node distance = 3cm] (balance2) {\small$\G^{\N^3(\calZ_1)} \to \G^{\N^2(\calZ_1)}$\normalsize \\ balance violated?};
\node [cloud,right of =myhillbalance1, node distance = 7cm] (non3){Non-surjective}; 	   
\node [cloud, below of =non3, node distance = 2cm] (non4) {Non-surjective};
\node [cloudunknown, below of =non4, node distance = 2cm] (unknown) {Unknown};

% Draw edges
\path [line] (init) -- (ruletable);
\path [line] (ruletable) -- node [midway] {no} (non1);
\path [line] (ruletable) -- node [midway] {yes} (depend);
\path [line] (depend) -- node [midway] {no} (non2);
\path [line] (depend) -- node [midway] {yes} (slice);
\path [line] (slice) -- node [midway] {yes} (good1);
\path [line] (slice) -- node [midway] {no} (myhillbalance1);
\path [line] (myhillbalance1) -- node [midway] {no} (balance2);
\path [line] (myhillbalance1) -- node [midway] {yes} (non3);
\path [line] (balance2) -- node [near end] {no} (non4);
\path [line] (balance2) -- node [midway,below] {yes} (unknown);
\end{tikzpicture}
\caption{Surjectivity test flowchart.}
\label{fig:flowchart}
\end{centering}  
\end{figure*}
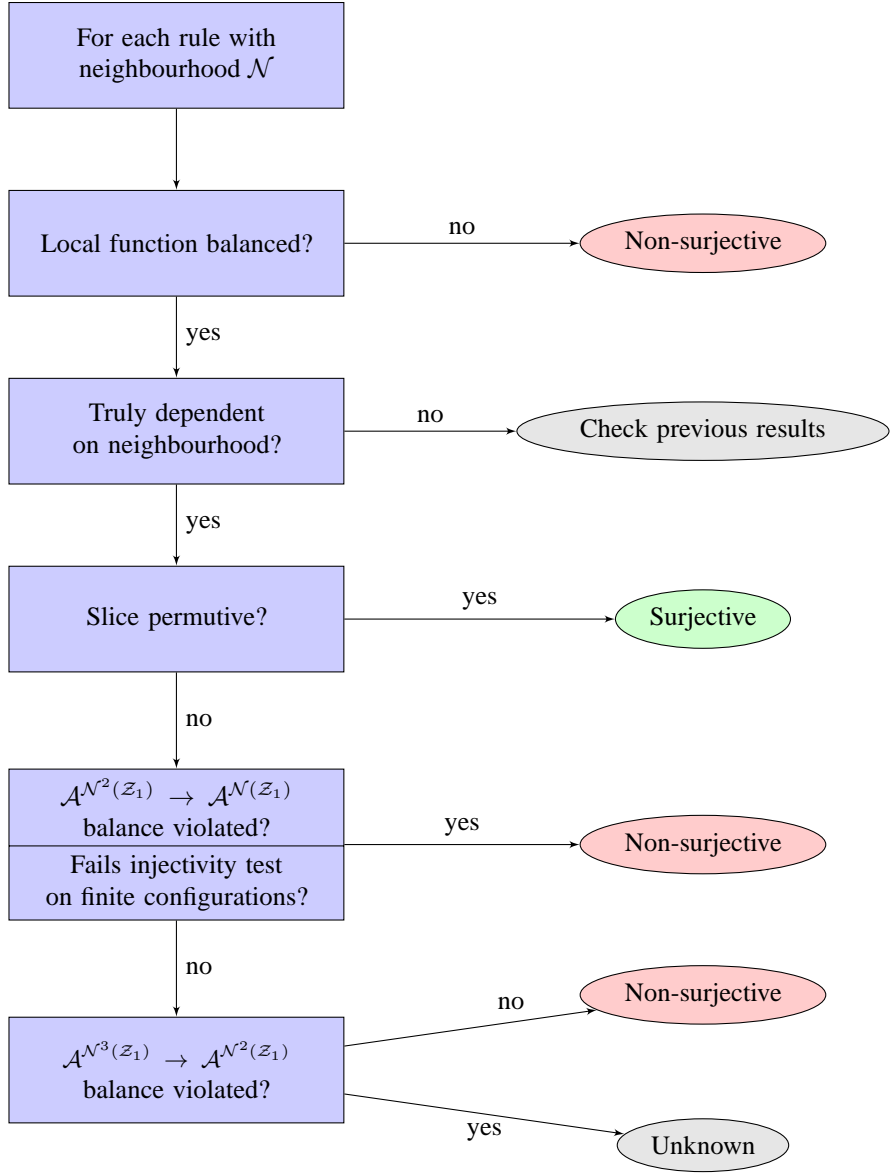

\begin{figure*}
\centering
\subfloat[Three-site]{\label{fig:threesite}
\begin{tikzpicture}
%\draw[ densely dashed] (0,0) circle (3pt);
%\draw[ densely dashed] (0.3,0) circle (3pt);
%\draw[ densely dashed] (-0.3,0) circle (3pt);
\draw[ densely dashed] (1.3,0) circle (3pt);
\draw[ densely dashed] (1.3,0.3) circle (3pt);
\draw[ densely dashed] (1.6,0) circle (3pt);
\draw[ densely dashed,white] (2.4,0) circle (3pt);
\draw[ densely dashed,white] (0.5,0) circle (3pt);
%\draw (0,-0.6)[] node{I};
\draw (1.45,-0.6)[] node{L};
\draw (1,-0.75)[white] node{$\star$};
\end{tikzpicture}} \hspace{20mm}
\subfloat[Four-site]{\label{fig:foursite}
\begin{tikzpicture}
%\draw[ densely dashed] (0,0) circle (3pt);
%\draw[ densely dashed] (0.3,0) circle (3pt);
%\draw[ densely dashed] (-0.3,0) circle (3pt);
%\draw[ densely dashed] (0.6,0) circle (3pt);
%\draw (0.15,-0.6)[] node{I};
\draw[ densely dashed] (1.5,0) circle (3pt);
\draw[ densely dashed] (1.5,0.3) circle (3pt);
\draw[ densely dashed] (1.8,0) circle (3pt);
\draw[ densely dashed] (2.1,0) circle (3pt);
\draw (1.8,-0.6)[] node{L};
\draw[ densely dashed] (3,0) circle (3pt);
\draw[ densely dashed] (3,0.3) circle (3pt);
\draw[ densely dashed] (3.3,0) circle (3pt);
\draw[ densely dashed] (3.3,0.3) circle (3pt);
\draw (3.15,-0.6)[] node{O};
\draw[ densely dashed] (4.5,0.3) circle (3pt);
\draw[ densely dashed] (4.5,0.6) circle (3pt);
\draw[ densely dashed] (4.8,0.3) circle (3pt);
\draw[ densely dashed] (4.8,0) circle (3pt);
\draw (4.65,-0.6)[] node{S};
\draw[ densely dashed] (6,0) circle (3pt);
\draw[ densely dashed] (6.3,0) circle (3pt);
\draw[ densely dashed] (6.3,0.3) circle (3pt);
\draw[ densely dashed] (6.6,0) circle (3pt);
\draw (6.3,-0.6)[] node{T};
\draw (3,-0.75)[white] node{$\star$};
\end{tikzpicture}}

\vspace{3mm}
\subfloat[Five-site]{\label{fig:fivesite}
\begin{tikzpicture}
\draw[ densely dashed] (2.1,0) circle (3pt);
\draw[ densely dashed] (1.8,0.3) circle (3pt);
\draw[ densely dashed] (1.8,0.6) circle (3pt);
\draw[ densely dashed] (2.1,0.3) circle (3pt);
\draw[ densely dashed] (2.4,0.3) circle (3pt);
\draw (2.1,-0.6)[] node{F};
%\draw[ densely dashed] (1.3,0) circle (3pt);
%\draw[ densely dashed] (1.6,0) circle (3pt);
%\draw[ densely dashed] (1.9,0) circle (3pt);
%\draw[ densely dashed] (2.2,0) circle (3pt);
%\draw[ densely dashed] (2.5,0) circle (3pt);
%\draw (1.9,-0.6)[] node{I};
\draw[ densely dashed] (3.2,0) circle (3pt);
\draw[ densely dashed] (3.2,0.3) circle (3pt);
\draw[ densely dashed] (3.2,0.6) circle (3pt);
\draw[ densely dashed] (3.2,0.9) circle (3pt);
\draw[ densely dashed] (3.5,0) circle (3pt);
\draw (3.35,-0.6)[] node{L};
\draw[ densely dashed] (4.2,0) circle (3pt);
\draw[ densely dashed] (4.5,0) circle (3pt);
\draw[ densely dashed] (4.2,0.3) circle (3pt);
\draw[ densely dashed] (4.5,0.3) circle (3pt);
\draw[ densely dashed] (4.2,0.6) circle (3pt);
\draw (4.35,-0.6)[] node{P};
\draw[ densely dashed] (5.2,0.3) circle (3pt);
\draw[ densely dashed] (5.5,0.3) circle (3pt);
\draw[ densely dashed] (5.5,0) circle (3pt);
\draw[ densely dashed] (5.8,0) circle (3pt);
\draw[ densely dashed] (6.1,0) circle (3pt);
\draw (5.65,-0.6)[] node{S};
\draw[ densely dashed] (6.8,0) circle (3pt);
\draw[ densely dashed] (7.1,0) circle (3pt);
\draw[ densely dashed] (7.4,0) circle (3pt);
\draw[ densely dashed] (7.1,0.3) circle (3pt);
\draw[ densely dashed] (7.1,0.6) circle (3pt);
\draw (7.1,-0.6)[] node{T};
\draw[ densely dashed] (8.1,0) circle (3pt);
\draw[ densely dashed] (8.4,0) circle (3pt);
\draw[ densely dashed] (8.7,0) circle (3pt);
\draw[ densely dashed] (8.1,0.3) circle (3pt);
\draw[ densely dashed] (8.7,0.3) circle (3pt);
\draw (8.4,-0.6)[] node{U};
\draw[ densely dashed] (9.4,0) circle (3pt);
\draw[ densely dashed] (9.7,0) circle (3pt);
\draw[ densely dashed] (10,0) circle (3pt);
\draw[ densely dashed] (9.4,0.3) circle (3pt);
\draw[ densely dashed] (9.4,0.6) circle (3pt);
\draw (9.7,-0.6)[] node{V};
\draw[ densely dashed] (10.7,0.3) circle (3pt);
\draw[ densely dashed] (10.7,0.6) circle (3pt);
\draw[ densely dashed] (11,0.3) circle (3pt);
\draw[ densely dashed] (11,0) circle (3pt);
\draw[ densely dashed] (11.3,0) circle (3pt);
\draw (11,-0.6)[] node{W};
\draw[ densely dashed] (12,0.3) circle (3pt);
\draw[ densely dashed] (12.3,0.3) circle (3pt);
\draw[ densely dashed] (12.6,0.3) circle (3pt);
\draw[ densely dashed] (12.3,0) circle (3pt);
\draw[ densely dashed] (12.3,0.6) circle (3pt);
\draw (12.3,-0.6)[] node{X};
\draw[ densely dashed] (13.3,0) circle (3pt);
\draw[ densely dashed] (13.6,0) circle (3pt);
\draw[ densely dashed] (13.9,0) circle (3pt);
\draw[ densely dashed] (14.2,0) circle (3pt);
\draw[ densely dashed] (13.9,0.3) circle (3pt);
\draw (13.75,-0.6)[] node{Y};
\draw[ densely dashed] (14.9,0.3) circle (3pt);
\draw[ densely dashed] (14.9,0.6) circle (3pt);
\draw[ densely dashed] (15.2,0.3) circle (3pt);
\draw[ densely dashed] (15.5,0.3) circle (3pt);
\draw[ densely dashed] (15.5,0) circle (3pt);
\draw (15.2,-0.6)[] node{Z};
\draw (3,-0.75)[white] node{$\star$};
\end{tikzpicture}}
%\end{centering}
\caption{Contiguous Neighbourhoods}
\label{fig:neighbourhoods}
\end{figure*}
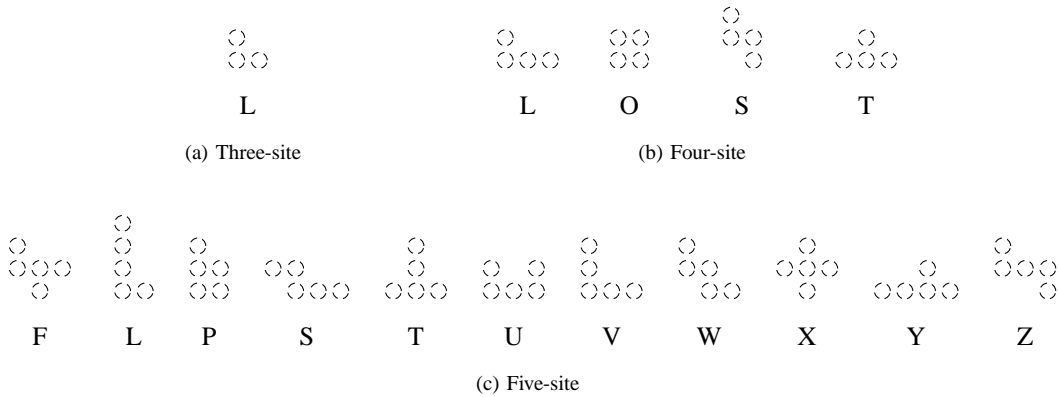 

\section{Classification Results}
We applied the aforementioned procedure to two-dimensional rules with neighbourhoods of size 3, 4, and 5.
It is not necessary to consider smaller neighbourhoods, as those are effectively one-dimensional, thus
 results of sec. \ref{results1d} apply. In all cases we considered only truly two-dimensional contiguous neighbourhoods,
meaning that linear neighbourhoods were excluded -- again, these are included in results of sec. \ref{results1d}.
Contiguous neighbourhoods have shapes known as polyominoes, that is, plane geometric figures formed by joining several equal squares edge to edge. Since  rigid transformations  of
$\ZZ^2$ preserve surjectivity, we considered only shapes which are representatives of equivalence
classes with respect to the group of isometries of $\ZZ^2$.

\subsection{Three-site neighbourhoods}
There is only one contiguous three-site neighbourhoods which is truly two-dimensional: the L-shaped neighbourhood (Figure \ref{fig:threesite}), with 256 corresponding binary rules.
Applying the procedure of  Figure \ref{fig:flowchart}, we found that all these rules can be classified, and that among them 38 rules are surjective, all of which are slice permutive. The remaining rules are non-surjective.

\vspace{4mm}

\begin{prop}
Any contiguous three-site binary CA is surjective if and only if it is slice permutive.
\end{prop}

\vspace{2mm}

%%%%%%%%%%%

\subsection{Four-site neighbourhoods}
There are four contiguous two-dimensional four-site neighbourhoods, often referred to as {\em tetrominos} and named after their resemblance to the letters L, O, S and T (Figure \ref{fig:foursite}). There are 65536 total binary rules dependent on each of these neighbourhoods. 

In the case of the L and T neighbourhoods, there are 724 slice permutive (and thus surjective) rules. In the case of the O and S neighbourhoods, there are 942 slice permutive and surjective rules. This difference seems to be due  to the number of sliceable sites in each neighbourhood (3 and 4 respectively). In each case, following Procedure \ref{proc:surjectivity} we determined that these are the only surjective rules, thus we again obtain a complete classification of all rules truly dependent on all four sites. This can be summarized as follows.

\vspace{5mm}

\begin{prop}
Any contiguous four-site two-dimensional binary cellular automaton is surjective if and only if it is slice permutive.
\end{prop}
\vspace{3mm}

\subsection{Five-site neighbourhoods}
There are eleven contiguous two-dimensional five-site neighbourhoods, often referred to as {\em pentominos} and named after their resemblance to the letters F, L, P, S, T, U, V, W, Z, Y and Z. There are 4294967296 binary rules associated with each neighbourhood. The results of application of Procedure \ref{proc:surjectivity} are shown in Table \ref{table:2Dfivesite}.
%%%%%%%%%%    
\begin{table*}
\centering
\caption{Two-dimensional truly five-site binary rules}\label{table:2Dfivesite}
\begin{tabular}{c|c|c||c|c|c}
Neighbourhood & Surjective & Unknown & Neighbourhood & Surjective & Unknown \\ \hline\hline
F & 257106 & 16 & V & 193138 & 0\\ \hline
L & 193138 & 1472 & W & 257106 & 3596\\ \hline
P & 257106 & 3596 & X & 257106 & 0\\  \hline
S & 192938 & 67764 & Y &193138 & 1472\\  \hline
T & 193138 & 0 & Z & 257106 & 0\\ \hline
U & 192938 & 64264 
\end{tabular} 
\end{table*}
We can observe that the classification is complete in the case of four neighbourhood shapes, T, V, X, and Z.

\vspace{6mm}

\begin{prop}
Any contiguous five-site two-dimensional binary cellular automaton with the neighbourhood
shaped as T, V, X, or Z  pentomino is surjective if and only if it is slice permutive.
\end{prop}

\vspace{6mm}

Note that this includes the traditional von Neumann neighbourhood (shape X).  Also note that for neighbourhood shapes
for which complete classification was not possible, the vast majority of rules were nevertheless classified.
In all cases the fraction of rules which were classified exceeded 99.998\%.

\section{Conclusions}
We were able to obtain complete classification with respect to surjectivity of all 2D rules with
contiguous neighbourhoods of size up to 4. In these neighbourhoods, all surjective rules are slice
permutive. Among 5-site rules, those with von Neuman neighbourhoods as well as neighbourhoods corresponding
to T, V, and Z pentominos can also be completely classified, and again, surjectivity and slice permutivity are
equivalent for them. For the remaining pentomino shapes, only a very small fraction of rules defies classification.
The worst case is the S pentomino, for which .0016\% rules cannot be classified with our algorithm.

A number of open questions remain. First of all, can the remaining 5-site rules be classified?
We suspect that at least some of them could be, if one performed balance and/or injectivity tests
on larger blocks, although such tests would be computationally very expensive. A related question
is whether there exist any truly two-dimensional five-site binary rule which is surjective yet not slice permutive? Since
in one dimension such rules are possible even in four-site neighbourhoods, we suspect that the answer is
affirmative, although currently we cannot offer any evidence of this claim.

\section{Acknowledgements}
One of the authors (HF) acknowledges financial support from the Natural
Sciences and Engineering Research Council of Canada (NSERC) in the
form of Discovery Grant. This work was made possible by the facilities of the Shared
Hierarchical Academic Research Computing Network (SHARCNET:www.sharcnet.ca) and
Compute/Calcul Canada. 

\bibliography{biblio}
\bibliographystyle{IEEEtran}

\end{document}